# FLANNEL: Focal Loss Based Neural Network Ensemble for COVID-19 Detection

**Zhi Qiao [1], Austin Bae[1], Lucas M. Glass[1], Cao Xiao [1], and Jimeng Sun[2]**
[1]Analytics Center of Excellence, IQVIA, Cambridge United States and Beijing, China,
[2]Computer Science Department, University of Illinois at Urbana-Champaign, Urbana, United States.

Correspondence to Cao Xiao, IQVIA, 201 Broadway Floor 5, Cambridge MA 02139, USA; Email: cao.xiao@iqvia.com

# Abstract

**Objective:** To test the possibility of differentiating chest x-ray images of COVID-19 against other pneumonia and healthy patients using deep neural networks.

**Materials and Methods:** We construct the X-ray imaging data from two publicly available sources, which include 5508 chest x-ray images across 2874 patients with four classes: normal, bacterial pneumonia, non-COVID-19 viral pneumonia, and COVID-19. To identify COVID-19, we propose a Focal Loss Based Neural Ensemble Network (FLANNEL), a flexible module to ensemble several convolutional neural network (CNN) models and fuse with a focal loss for accurate COVID-19 detection on class imbalance data.

**Results:** FLANNEL consistently outperforms baseline models on COVID-19 identification task in all metrics. Compared with the best baseline, FLANNEL shows a higher macro-F1 score with 6% relative increase on Covid-19 identification task where it achieves 0.7833±0.07 in Precision, 0.8609±0.03 in Recall, and 0.8168±0.03 F1 score.

**Discussion:** Ensemble learning that combines multiple independent basis classifiers can increase the robustness and accuracy. We propose Neural Weighing Module to learn importance weight for each base model and combine them via weighted ensemble to get the final classification results. In order to handle the class imbalance challenge, we adapt Focal loss to our multiple classification task as the loss function.

**Conclusion:** FLANNEL effectively combines state-of-the-art CNN classification models and tackle class imbalance with Focal loss to achieve better performance on Covid-19 detection from X-rays.

# I. OBJECTIVE

The novel coronavirus disease (COVID-19) is a pandemic disease that has been spreading rapidly across the world and poses a serious threat to global public health. Till July 2020, COVID-19 has caused more than 13 million infections and 600k deaths across the globe. Radiology plays a fundamental role in diagnosing the disease. Although less sensitive than chest CT, chest radiography (X-ray) is typically the first-line imaging modality used for patients with suspected COVID-19 [1]. Due to the international attention of the disease, there has been an increase in publicly available X-ray images of patients with COVID-19 related Pneumonia. This enables us to identify patterns and construct models that could automatically detect COVID-19 from chest X-ray images.

Because COVID-19 is still relatively new, naturally there are significantly less X-ray images of COVID-19 compared to non-COVID-19 diseases. This inevitably means that these datasets will be imbalanced with few COVID-19 samples which can hinder classification accuracy and limit model generalization.

To address these challenges, we propose a Focal Loss bAsed Neural Network EnsembLe (FLANNEL) model. FLANNEL utilizes an ensemble structure, using five state-of-the-art convolutional neural network (CNN) classifiers as based models. The Neural Weight Module combines each base model using a learnt weighted ensemble to get the final classification results. To handle the class-imbalanced dataset issue, the model adapts Focal loss traditionally proposed for binary classification task to our multi-class classification task. This loss function back-propagates and updates the parameters of the Neural Weight Module.

We train and evaluate the proposed FLANNEL architecture on a combination of publicly available datasets. The combined data consists of 5508 chest x-ray images across 2874 patients. There are 4 types of X-ray images: Normal (n=1118), Bacterial Pneumonia

(n=2787), COVID-19 Pneumonia (n=100), and non-COVID-19 viral Pneumonia (n=1503). These 4 types of X-rays naturally form a 4-way classification problem, where our focus is to make accurate prediction on COVID-19 cases without hindering the performance of other classes.

## II. BACKGROUND AND SIGNIFICANCE

### Convolutional Neural Networks

The convolutional neural networks (CNNs) have established themselves as the golden standard for most computer vision tasks such as image classification and object detection. The basic building blocks for all CNN architectures are convolutional layers that extract feature maps from the input followed by pooling layers and a fully-connected and softmax layer at the very end. The yearly The ImageNet Large Scale Visual Recognition Challenge (ILSVRC) competition has motivated some of the best state-of-the-art CNN architectures, showing improvements on image classification tasks on standard image datasets such as CIFAR[1] and ImageNET[2] every year. Starting from standard 8-layer AlexNet [2], CNNs have gotten more sophisticated with the addition of techniques such as Inception Modules and Residual Blocks [3-6].

For specific COVID-19 challenge, some works based on CNNs are created for various imaging data. Most of them focus on region segmentation and COVID diagnosis [21-31]. For example, [21] designed and developed a contactless patient positioning system for scanning patients in a completely remote and contactless fashion. Chest CT based COVID-19 cases identification or localization segmentation tasks are studied in [22-30]. [31] presented a deep learning method for COVID-19 classification based on lung ultrasound data. Several of them make prediction on 3D images (sequential CT-slice / Ultrasound-slice) [22,24,25,26,27,29,30,31]. Different from those existing works, we focus on identification task on X-ray images which are 2D images as they are more commonly used in routine clinical visits. As the first-line imaging modality for assessing patients with suspected COVID-19, 2D chest X-ray is less sensitive than chest CT and hence posts a bigger technical challenge. The most related work is COVID-Net, a neural network architecture for X-ray based COVID cases detection [20]. We added COVID-Net [20] and AI-COVID [22] as baselines in the experiments.

---

[1] https://www.cs.toronto.edu/~kriz/cifar.html
[2] http://www.image-net.org/

Moreover we adopted several state-of-the-art CNN architectures as base models in FLANNEL for X-ray based COVID-19 detection. Due to limited number of COVID-19 X-ray images exist in training data, each of these models alone could suffer from high error and variance. In order to solve this issue and improve reliability, we utilize the power of Ensemble methods.

## Ensemble Models

Ensemble Learning is a process of combining different independent classifiers, called base learners, to increase the robustness and accuracy of the final classification. As long as these base learners are diverse and able to capture the patterns in the data independently, the ensemble model is able to generalize significantly better by reducing individual errors [7]. Instead of training a complex single neural network with a large number of layers and parameters, decomposing the architecture into smaller and simpler individual base models have shown to be more accurate and require less time and memory for training [8,9]. The individual base learners in Neural Network Ensembles capture patterns over different regions and/or granularities of the input space [10], which contributes to its higher overall performance. Instead of relying on traditional methods of Ensemble voting such as Bagging, our proposed model introduces a Neural Ensemble Layer to adopt a Heterogeneous Ensemble strategy that uses different state-of-the-art CNNs as base learners to get the better accuracy compared with the individual classifiers.

## Class Imbalance Challenge

Class imbalance in datasets is a very common but significant problem that often hinders model performance and generalization. Re-sampling, via undersampling or oversampling is a common tactic to solve the imbalance issue [11-13]. However, oversampling tends to be error-prone due to overfitting or added noise and undersampling reduces the amount of training data the model can learn from [14], causing both methods ineffective in our setting. GANs (Generative Adversarial Networks) have shown to be decent augmentation tools to rebalance such datasets [15], but GANs are quite difficult to train and can lead to the same problems as oversampling. Focal Loss is a special loss function proposed for imbalance binary classification task, in which the standard Cross-Entropy loss of model is reshaped such that the well-classified examples are down-weighted and it can focus on learning the hard imbalanced negatives [16]. In this paper, we adapt Focal loss to our multiple classification task.

# III. Materials and Methods

## Study design

Our main task is to differentiate between chest x-ray images of COVID-19 and other pneumonia and healthy patients which can be considered as a multiple classification task.

We combined two publicly available datasets: 1) COVID Chest X-ray (CCX) dataset[3] and 2) Kaggle Chest X-ray (KCX) images (Pneumonia) dataset[4] to generate the experimental data. The Kaggle chest X-ray data only contains three types of X-ray images: normal, Bacterial Pneumonia, and non-COVID-19 viral Pneumonia. We get the COVID-19 Pneumonia images from the CCX dataset, which contains a few X-ray images from the other classes as well. The choice of these two datasets for our experiment is guided by the fact that both are open source and fully accessible to the research community and the general public. As these public datasets grow, the model can continue to improve and scale continuously.

---

[3] https://github.com/ieee8023/covid-chestxray-dataset
[4] https://www.kaggle.com/paultimothymooney/chest-xray-pneumonia

All of KCX images and most CCX images are based on anteroposterior (AP) or posteroanterior (PA) view. For consistency purposes, we only included images with both AP and PA view in our final experimental dataset. The dataset comprises a total of 5508 chest radiography images across 2874 independent patient cases. All images in KCX data are all in AP/PA view. CCX data has 131 images, of which 119 are AP/PA images. We only selected the 119 AP/PA images in our experiments, which includes 100 covid-related pneumonia, 11 viral pneumonia, 7 bacterial pneumonia, and 1 normal patient Chest X-ray images.

Because AP and PA are two different type of X-ray views, we introduced horizontal flips and random noise to convert PA into AP view. This method of data preprocessing allows us to train the same model for both views and allows the model to be view-independent. We used a train-test split ratio of 4:1 to randomly generate the training and testing set to be used in the model. We furthermore use 5-fold cross-validation on training data to acquire 5 resulting models, and then record ensemble (average) performance of these 5 models on testing data. The reason we did this instead of one-time standard train, validation, test splits is that we want to maximize the sample usage because of the limited samples in this study. The detailed statistics of experimental data are shown in **Table 1**.

For image preprocessing, we resized the original input images first to size 256x256 and then randomly cropped them in the center to size 224x224. The original X-ray images have some symbols in the corners correlated with our labels that could lead to feature leakage, so we introduce this crop around the center to mask that. The resolution 224x224 was chosen as it is the default input size for all of our base models. Although this downsampling in resolution may lead to loss of image details, it allows a simpler model that is easier to train and less prone to overfitting. Most state-of-the-art CNNs adopt this resolution for the same reason. The data is then augmented using random flip and noise and normalized before entering the model.

Table 1 Experimental Data Description

| Source | | Total | COVID-19 | Viral | Bacterial | Normal |
|---|---|---|---|---|---|---|
| Original Data | CCX data | 119 | 100 | 11 | 7 | 1 |
| | KCX data | 5389 | 0 | 1492 | 2780 | 1117 |
| View Distribution | AP View | 5413 | 24 | 1492 | 2780 | 1117 |
| | PA View | 95 | 76 | 11 | 7 | 1 |
| Training/Test splits | Training | 4406 | 77 | 1194 | 2237 | 898 |
| | Testing | 1102 | 23 | 309 | 550 | 220 |
| | Total | 5508 | 100 | 1503 | 2787 | 1118 |

From **Table 1**, we can find that COVID-19 X-ray images are very rare compared to that of non-COVID-19 Pneumonia, because it is such a new disease with limited data.

## Methods

As shown in **Figure 1**, our proposed FLANNEL is composed of two stages. First, several base learners are independently trained for COVID-19 classification. Second, the resulting weights from all base learners are used to train the overall ensemble model. We also provide pseudocode in **Table 2** to elaborate on the algorithm in detail.

**Stage-1 Base Learner Training**

The convolutional neural networks (CNNs) have been widely used in image classification and get huge successes. Here, we choose 5 popular and state-of-the-art CNN classification models as base learners to model the COVID-19 identification task. These following models were chosen due to their flexibility and high performance with general image classification.

1. Inception v3: It is the third edition of Google's Inception Convolutional Neural Network [3,17]. This model utilizes Inception Modules inspired by Network-in-Networks [4].
2. VGG19-bn: The model architecture is from VGG group with batch normalization and consists of 19 layers, where 16 Convolutional layers and 3 Fully Connected layers.
3. ResNeXt101: This 101-layer architecture is designed by the ResNeXt group.
4. Resnet152: This is a 152-layer Deep Residual Neural Network that learns the residual representation functions instead of directly learning the signal representations.
5. Densenet161: This is a Densely Connected Convolutional Network with 161 layers [18].

Due to the limited amount of training data of X-ray images, we use the listed pretrained models from the ImageNet Large Scale Visual Recognition Challenge (ILSVRC[5]) and fine-tune each model with respect to the COVID-19 identification task. For fine-tuning, all parameters of the models were retrained with no layer being frozen. This method of fine-tuning models pretrained on a general image set speeds up the training process and also helps with generalization. We modify the last classification layer for each base learner such that it produces a 4-length vector for 4-way classification.

**Stage-2 Ensemble Model Learning**

We then take all the N (number of pretrained base learners and here is 5) $\mathbb{M}$-length vectors (denoted as $P_i \in \mathbb{R}^M$, $i = 1, \ldots, N$, where $\mathbb{R}$ represents a real number) and simply concatenate them (denoted as $f$) and feed it into the **Neural Weight Module** to learn base learner weights as shown in **Figure 1**.

---

[5] http://www.image-net.org/challenges/LSVRC/

For the Neural weight Module, we firstly construct feature interaction via outer production $ff^T$ to capture more latent information. This is then flattened and fed through a Dense and Tanh layer to map features into base learner weights. Because the outputs of the Tanh function can output negative values, such weights allow the model to discount inaccurate predictions of the classifier.

We then take the linear combination of the base learner predictions using the resulting base learner weights (denoted as *w*) to get final prediction $\hat{y} = Softmax(\sum_{i=1}^{N} w_i P_i)$. Rather than relying on traditional methods of Ensemble voting, we allow FLANNEL to self-learn the optimal combination between the output of the base learners. The neural weight module could be easily extended to accommodate more complex output formats of the base learners.

Then we define loss function for model training. The standard loss function used for training multiclass Neural Networks is Cross-Entropy Loss as following,

$$LossFunc = CELoss(\hat{y}, y) = \sum_{m=1}^{M} -y_m log(\hat{y}_m)$$

where $M$ represents the number of classes. Both $\hat{y} \in R^M$ and $y \in \{0, 1\}^M$ are $M$-length vectors and represent the predicted value and ground truth of class distribution, respectively.

However, heavy class imbalances in the data during training will overwhelm and dominate the gradient, making optimal parameter updates difficult. **Focal Loss** is a loss function proposed for binary classification tasks, where the well-classified examples are down-weighted and can focus on learning the hard imbalanced examples [16]. Here, we extend focal loss to multiclass classification in our model to address these imbalance issues. For each image, we define the focal loss as:

$$LossFunc = FocalLoss(\hat{y}, y) = \sum_{m=1}^{M} -\alpha_m y_m (1 - \hat{y}_m)^\gamma log(\hat{y}_m)$$

where $(1 - \hat{y}_m)^\gamma$ is a modulating factor with a tunable focusing parameter $\gamma$, and $\alpha_m$

represents a weight factor vector that balances the importance of the different classes. When an example is misclassified and $\hat{y}_m$ is small, the modulating factor is close to 1 and the loss is unaffected. As $\hat{y}_m \to 1$, the factor goes to 0 and the loss for well-classified examples is down-weighted. $\gamma$ can be adjusted to tune the rate of this down-weighting. As increases, the effect of the modulating factor is likewise increased (we found $\gamma = 3$ to work best in our experiments). For $\alpha_m$, we can set the bigger value for the minority class and the smaller value for the majority class to make all classes contribute equally to our loss. When $\{\alpha_m = 1, m = 1 \ldots M\}$ and $\gamma = 0$, Focal Loss is equivalent to cross entropy loss. In our experiments, $\alpha_m$ is set to be inverse class frequency of each class. The resulting loss is then back-propagated to update the weights for the Neural Weighing module in the ensemble. During this stage, the parameters of the five base learners are frozen and not updated.

Table 2 The FLANNEL Algorithm

| **Algorithm 1** FLANNEL Training |
|---|
| **Input:** <br> X-ray Images, Class Labels <br> Base Models {Learner$_1$, Learner$_2$, …, Learner$_n$} <br> *(Define B as batch size)* <br> **Stage 1:** <br> Fine-tune all base models with respect to inputs images and labels. <br> **Stage 2 :** <br> **For** each batch $(X \in \mathrm{R}^{B \times 1 \times 224 \times 224}, Y \in \mathrm{R}^{B \times 4})$ from inputs images and labels do <br>    --> *Step1: Get prediction values from all Base Models* <br>    $P_i = Learner_i(X) \in \mathrm{R}^{B \times 4}$, where $i = 1, \ldots, n$ <br>    --> *Step2: Get learner weights* <br>    $W = NeuralWeightModule([P_i,\ i = 1, \ldots, n]) \in \mathrm{R}^{B \times 5}$ <br>    --> *Step3: Linear Combination for Prediction* <br>    $\hat{Y} = Softmax(\sum_{i=1}^{n} W_i P_i) \in \mathrm{R}^{B \times 4}$ <br>    (where $W_i$ represents i-th column of $W$) <br>    --> *Step4:* <br>    Loss = $FocalLoss(\hat{Y}, Y)$ <br>    Back-propagate on Loss and update parameters <br> End For |

## IV. RESULTS

### Baseline models for performance comparison

Firstly, we compare FLANNEL with the chosen 5 base learners of FLANNEL framework,

Resnet152, Densenet161, InceptionV3, VGG19-bn, and ResNeXt101. These models are some of the most popular state-of-the-art CNNs for image classification, known to have high classification rates for most image datasets. All of these models were fine-tuned using their default parameter settings and by using the Adam optimizer [19].

Furthermore, we also compare FLANNEL with two recent Covid-19 deep learning models, COVID-Net[20] and AI-Covid[22]. COVID-Net is a tailored deep convolutional neural network design for COVID-19 cases detection. AI-Covid presented a deep network framework for sequential CT- slice image, and hence we just use slice-level classification networks for our identification task since we only have one X-ray image per sample in our dataset.

To verify the advantages of FLANNEL on ensemble learning, we also selected two traditional ensemble strategies voting and stacking to ensemble the 5 chosen base learners. These will be denoted as Voting and Ensemble_MLP respectively. For Ensemble_MLP, the concatenated prediction values from base models are fed into the Multi-Layer Perceptron for final prediction. Here, we use separately Ensemble_MLP_l1 for 1-layer MLP and Ensemble_MLP_l2 for 2-layer MLP.

To verify the advantages of FLANNEL for the imbalanced datasets, we compare it to a version of FLANNEL that replaces the Focal Loss with multiclass standard Cross-Entropy Loss (denoted as FLANNEL_w/o_focal). Another comparison model adds on top of FLANNEL_w/o_focal and utilizes resampling strategies commonly used for class balancing (denoted as FLANNEL_w/o_focal_sampling).

## Implementation details

All the base models and FLANNEL are implemented in PyTorch and trained on 3 NVIDIA Tesla P100 GPUs over 200 epochs. The 5 base models[6] are fine-tuned using the respective pre-trained models with the default model architecture. The data is augmented

---
[6] InceptionV3, Densenet161, Resnet152, ResNeXt101, Vgg19_bn

with random flips, crops and scaling during the fine-tuning process.

After all the base models are trained separately, FLANNEL is trained by passing in the concatenated output layers of the base models as the input features.

## Evaluation strategy

In order to overall verify the prediction accuracy, we firstly measure the overall accuracy of the model in distinguishing the four classes (COVID19 viral pneumonia, Non-COVID19 viral pneumonia, bacterial pneumonia and normal images). The main intention of the study is the detection of Covid-19 among kinds of respiratory related X-ray images. For each class of images, the classification metric F1-score, which conveys the balance between the precision and the recall, is recorded.

## Experimental Results

In this section, we present the experimental results to show the performance of our proposed FLANNEL and all baseline methods.

First, we note that overall accuracy is not a great metric for evaluation the model. Since the classes in our dataset is heavily unbalanced in favor of non-COVID-19 Pneumonia images, even a significant increase in COVID-19 detection performance will not affect overall accuracy very much. Therefore, we present the F1-score for COVID-19 vs. rest comparing different models in **Figure 2**. Obviously, it shows that FLANNEL outperforms other state-of-the-art models in detecting COVID-19 cases.

Moreover, we also present F1-score for each disease classification and macro-F1 score for all classes of the label set, which is shown in **Table 3**. From Table 3, we can observe that COVID-Net achieves better performance than other baseline models, since it is specifically designed for COVID detection. AI-Covid does not show apparent performance improvement on COVID detection. It is probably because AI-Covid mainly focus on fast features extraction followed by multi-slice ensemble for final 3-D CT image prediction, not specifically for 2D image classification. Multi-layer perceptron based ensemble strategy can improve detection performance on other majority classes, but has poor performance on minority class (COVID class). And, COVID-detection accuracy degrades with more-layers employment. More fully connected layers cannot effectively model limited features (concatenated outputs from base learners) specifically for this class-imbalance challenge. From **Table 3**, FLANNEL shows a higher macro-F1 score with 2% increase, especially a 6% relative increase for COVID-19 cases over the best performing baseline model.

Compared with FLANNEL_w/o_Focal just using Cross-Entropy Loss instead of Focal Loss, FLANNEL shows an increase of almost 6% in F1 score for COVID-19 cases. It is also clear that resampling strategies help improve performance in case of class imbalance. Traditional resampling strategies (FLANNEL_w/o_Focal + Sampling) increased F1 score for COVID-19 cases by almost 2%, while still almost 4% behind compared to FLANNEL. Also most importantly, FLANNEL was able to increase the performance of COVID-19 classification without negatively impacting the performance for other classes.

Table 3 Performance comparison on F1-score: Class-specific F1-score is calculated using one class vs. the rest strategy.

|  | Covid19 | Pneumonia Virus | Pneumonia Bacteria | Normal | Macro-F1 |
|---|---|---|---|---|---|
| **Base Learners** | | | | | |
| **InceptionV3** | 0.5904(0.27) | 0.5864(0.05) | 0.8056(0.01) | 0.8771(0.04) | 0.7149(0.09) |
| **Vgg19_bn** | 0.6160(0.06) | 0.5349(0.04) | 0.7967(0.02) | 0.8691(0.03) | 0.7042(0.02) |
| **ResNeXt101** | 0.6378(0.12) | 0.5649(0.03) | 0.7959(0.01) | 0.8537(0.02) | 0.7140(0.03) |
| **Resnet152** | 0.6277(0.11) | 0.5506(0.02) | 0.7988(0.01) | 0.8700(0.01) | 0.7110(0.03) |
| **Densenet161** | 0.6880(0.07) | 0.5930(0.02) | 0.8017(0.01) | 0.8953(0.01) | 0.7445(0.02) |
| **Additional Baselines** | | | | | |
| COVID-Net [20] | 0.7179(0.13) | 0.5592(0.04) | 0.8095(0.02) | 0.8787(0.03) | 0.7413(0.03) |
| AI-Covid [22] | 0.6391(0.16) | 0.5238(0.07) | 0.7504(0.02) | 0.7223(0.03) | 0.6589(0.07) |
| Ensemble Learning | | | | | |
| Voting | 0.7684(0.04) | 0.6005(0.03) | 0.8214(0.03) | 0.9079(0.01) | 0.7745(0.01) |
| Ensemble_MLP_l1 | 0.6247(0.07) | 0.6042(0.03) | 0.8185(0.01) | **0.9161(0.01)** | 0.7409(0.02) |
| Ensemble_MLP_l2 | 0.3735(0.23) | 0.6030(0.02) | 0.8206(0.00) | 0.9128(0.01) | 0.6775(0.05) |
| Variant FLANNEL | | | | | |
| FLANNEL_w/o_Focal | 0.7671(0.06) | 0.6001(0.03) | 0.8238(0.01) | 0.9135(0.01) | 0.7761(0.01) |
| FLANNEL_w/o_Focal + Sampling | 0.7837(0.04) | 0.5953(0.04) | 0.8245(0.01) | 0.9131(0.01) | 0.7791(0.02) |
| FLANNEL | **0.8168(0.03)** | **0.6063(0.02)** | **0.8267(0.00)** | 0.9144(0.01) | **0.7910(0.01)** |

In order to understand the trade-off in COVID-19 classification performance for different threshold values, we also introduce the plot curves. A popular metric for classification systems is the Receiving Operator Characteristic (ROC), which can be summarized by its area under the curve (AUC). Since ROC curves can be misleading when the class distribution is imbalanced, we also show the precision-recall curve (PR Curve). The focus of the PR curve on the minority class deems it an effective diagnostic for imbalanced binary classification models. The experimental results are shown in **Figure 3**, where PRC and ROC are separately used to show the diagnostic ability of compared models.

Finally, we also provide the visual description of FLANNEL performance, via Confusion Matrix, shown in **Figure 4**. It shows that 1) the cases predicted as pneumonia are mainly from the GroundTruth pneumonia related cases 2) for COVID-19 identification, FLANNEL has higher precision and recall than other two types of pneumonia. It means FLANNEL can distinguish pneumonia images from normal, and differentiate chest x-ray images of COVID-19 against other pneumonia images.

In order to verify model performance under single-view condition, we show performance evaluation (specifically for F1 score for COVID-19 vs. Rest) on AP-view X-ray images in **Figure 5**. From Figure 5, we can find that all models have performance degradation because of heavier imbalance with fewer samples with COVID label. Nevertheless, our proposed FLANNEL still has best F1 score compared to all the baselines.

Then, **Figure 6** presents the visual explanations for the classification results. We use Class Activation Mapping (CAM) to visualize the models' attention over inputs. The visual results are shown below. Here, we show each of the 5 base models' visual attention heatmaps. We observe that FLANNEL can pay more attention to specific regions in each base model parameters which allows it to make the correct overall prediction.

## V. Discussion

The most notable observation from the base models is that most of them perform poorly at detecting the Pneumonia caused by Viruses (both COVID-19 and non-COVID-19). This poor performance is likely due to the fact that 1) there are more bacterial and normal than viral images in our dataset and 2) the addition of the COVID-19 X-ray images makes it more difficult to differentiate between the two different classes of Viral Pneumonia since they are more similar to each other than they are to others. Our proposed FLANNEL has the possibility of differentiating chest x-ray images of COVID-19 against other pneumonia and healthy patients. Moreover, the significant feature of proposed FLANNEL is that it manages to sharply enhance COVID-19 identification performance without decreasing that of other classes. In fact, FLANNEL had the highest F1 score out of every single category except the normal class.

However, due to the limited specific COVID-19 Pneumonia related images, it is inevitable to combine multiple source data from different datasets to construct experimental data which has been described in section III in details. Even standard X-ray images look slightly different depending on the data source. For example, as noted before, there are some specific flags on images which originate from different X-ray acquisition equipment. The existing data source features can impact model learning for the core classification task. We have implemented some preprocessing methods, such as random center crop, to to alleviate the impacts from data sources. However, it is still a challenge to conduct source agnostic image classification that utilizes big scale data from multiple data source to learn class features and mapping rules to infer the classification task without considering the different data source impacts. Source-free classification for COVID-19 is future work that might significantly improve the performance of the model.

Another direction to consider is to incorporate hierarchical classification. In our case, we can consider first separating healthy from sick patients based on their X-rays, then further separating bacteria, viral pneumonia and COVID among the sick. The naïve approach requires training multiple models at different levels which can create challenges at the granular level (e.g., viral pneumonia vs. COVID) due to the small sample size. Multiple models also introduce more computation and maintenance challenges in the long run. A single model that directly incorporates the hierarchical class structure is an interesting but challenging task in itself. One can consider designing a proper loss function to capture the intricate class hierarchy. We consider this topic as an important future work.

## VI. CONCLUSION

COVID-19 is an acute resolved pneumonia disease with high fatality rate. As the most commonly ordered imaging study for patients with respiratory complaints, X-ray can help COVID-19 diagnosis. Hence, it is significant to study automatic diagnosis of the disease. Our objective is to explore effective deep learning methods to model X-ray images and improve prediction performance in identifying Covid-19 from other pneumonia images and healthy images.

With the power of Ensemble Learning, FLANNEL has the ability to detect and diagnose COVID-19 from Pneumonia X-ray images with high accuracy, even when trained on just around 100 available COVID-19 X-ray images. We have shown that it is able to automatically combine and use the outputs of individual base learners as features to create a more accurate global model. Focal Loss allows us to use all the training samples effectively without having to sample our already limited imbalanced dataset and solves the imbalance problem that hinders other traditional loss functions such as Cross-Entropy loss. FLANNEL vastly outperforms all other state-of-the art CNN architectures especially on the COVID-19 detection without much added model complexity or parameters.

This model could be used to supplement current COVID-19 diagnosis kits to improve testing availability and alleviate supply shortages through just examining X-ray images. With millions of confirmed cases only expected to grow in the future, neural networks could make a real impact in limiting the spread of the disease.

## FUNDING STATEMENT

This research received no specific grant from any funding agency in the public, commercial or not-for-profit sectors.

## COMPETING INTERESTS STATEMENT

The authors have no competing interests to declare.

## CONTRIBUTORSHIP STATEMENT

Zhi Qiao implemented the method and conducted the experiments. All authors were involved in developing the ideas and writing the paper.

COVID-19 markers in point-of-care lung ultrasound. *IEEE TMI 2020 Aug;39(8): 2676-2687.*

IMAGE LEGENDS

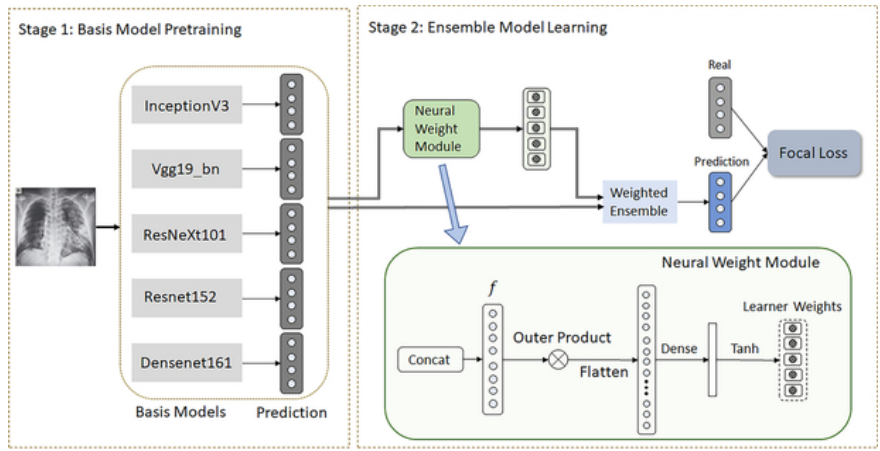

**Figure 1.** Framework of FLANNEL.

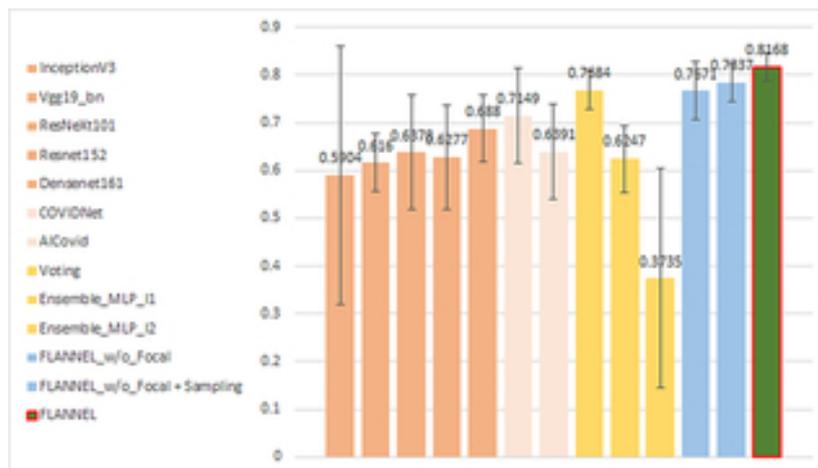

**Figure 2.** Illustrates the COVID-19 F1-score vs. rest comparing different models. (The error bars are from five-fold cross-validation)

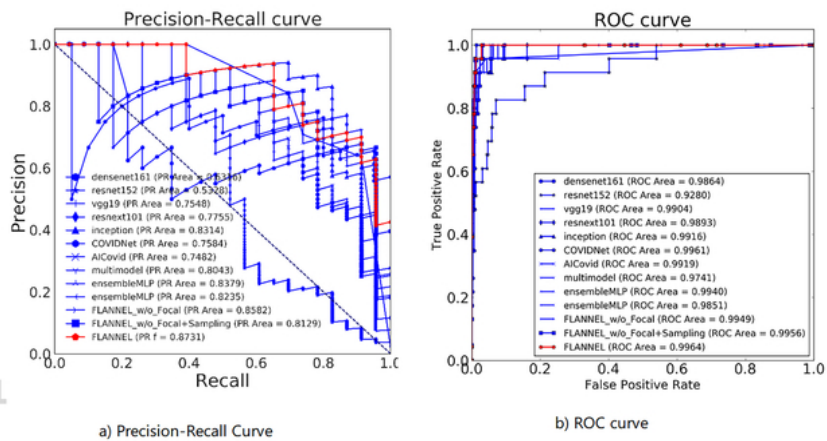

**Figure 3.** Illustrates the diagnostic ability of compared models for COVID-19 classification

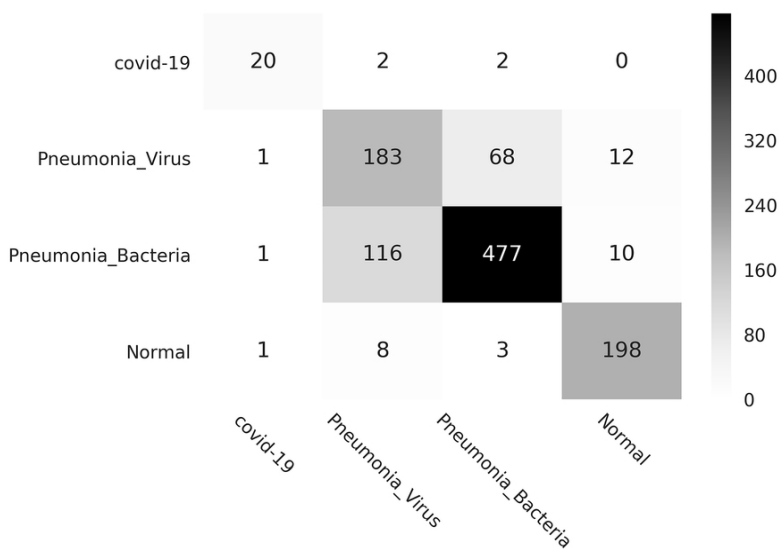

**Figure 4.** Confusion Matrix(instance counts)

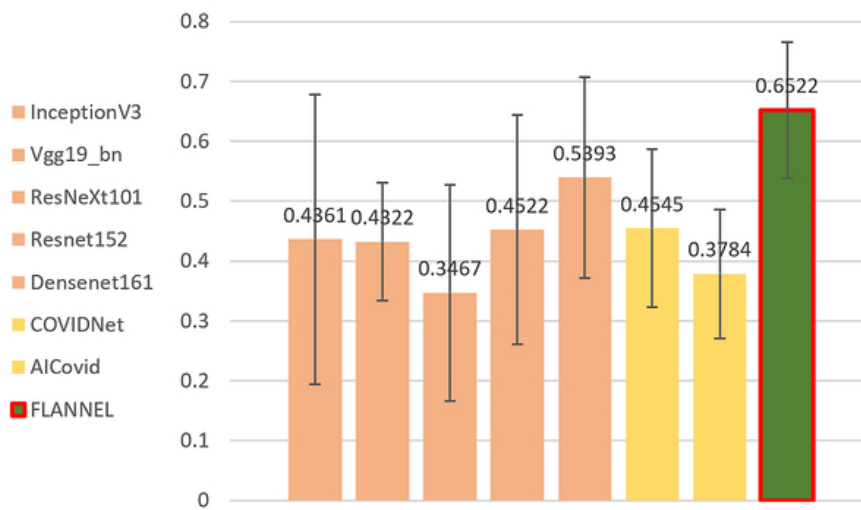

**Figure 5.** Performance verification (F1-score for COVID-19 vs. rest) on AP view images comparing different models.

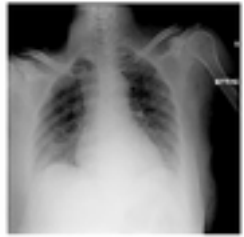

a) Original Covid X-ray

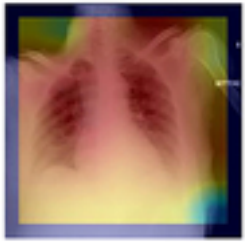 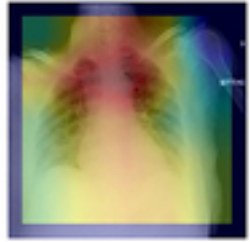 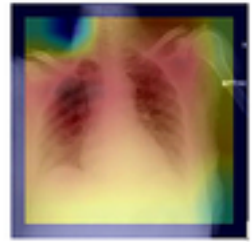

b) FLANNEL(P=0.976)  c) Inception_v3(P=0.817)  d) ResNext101_32x8d(P=0.801)

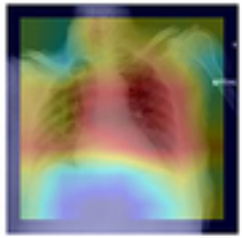 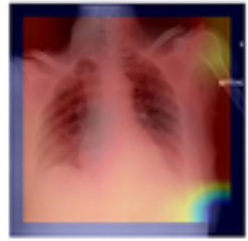 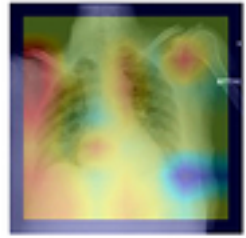

e) DenseNet161(P=0.001)  f) ResNet152(P=0.805)  g) Vgg19(P=0.442)

**Figure 6.** Visual explanations for the classification results